\documentclass[submission, Proceedings]{SciPost}

\hypersetup{
	colorlinks,
	linkcolor={red!50!black},
	citecolor={blue!50!black},
	urlcolor={blue!80!black}
}

\usepackage[bitstream-charter]{mathdesign}
\DeclareSymbolFont{usualmathcal}{OMS}{cmsy}{m}{n}
\DeclareSymbolFontAlphabet{\mathcal}{usualmathcal}

\begin{document}

\begin{center}{\Large \textbf{High-energy resummation in inclusive hadroproduction \\ of Higgs plus jet\\
}}\end{center}
 

\begin{center}
	F.~G.~Celiberto~$^{1,2,3}$,
	D.~Yu.~Ivanov~$^{4}$,
	M.~M.~A.~Mohammed~$^{5,*}$,
	A.~Papa~$^{5}$
\end{center}

\begin{center}
	
	\centerline{${}^1$ {\sl European Centre for Theoretical Studies in Nuclear Physics and Related Areas (ECT*),}}
    \centerline{\sl I-38123 Villazzano, Trento, Italy}
    \vskip .1cm
    \centerline{${}^2$ {\sl Fondazione Bruno Kessler (FBK), 
    I-38123 Povo, Trento, Italy} }
    \vskip .09cm
    \centerline{${}^3$ {\sl INFN-TIFPA Trento Institute of Fundamental Physics and Applications,}}
    \centerline{\sl I-38123 Povo, Trento, Italy}
	\vskip .09cm
	\centerline{${}^4$ {\sl Sobolev Institute of Mathematics, 630090 Novosibirsk,
			Russia}}
	\centerline{ {\sl and Novosibirsk State University, 630090 Novosibirsk,
			Russia}}
	\vskip .09cm
	\centerline{${}^5$ {\sl Dipartimento di Fisica, Universit\`a della Calabria,}}
	\centerline{{\sl and Istituto Nazionale di Fisica Nucleare, Gruppo collegato
			di Cosenza,}}
	\centerline{\sl I-87036 Arcavacata di Rende, Cosenza, Italy}
	\vskip .09cm
	* mohammed.maher@unical.it
\end{center}

\begin{center}
	\today
\end{center}

\definecolor{palegray}{gray}{0.95}
\begin{center}
\colorbox{palegray}{
  \begin{tabular}{rr}
  \begin{minipage}{0.1\textwidth}
    \includegraphics[width=22mm]{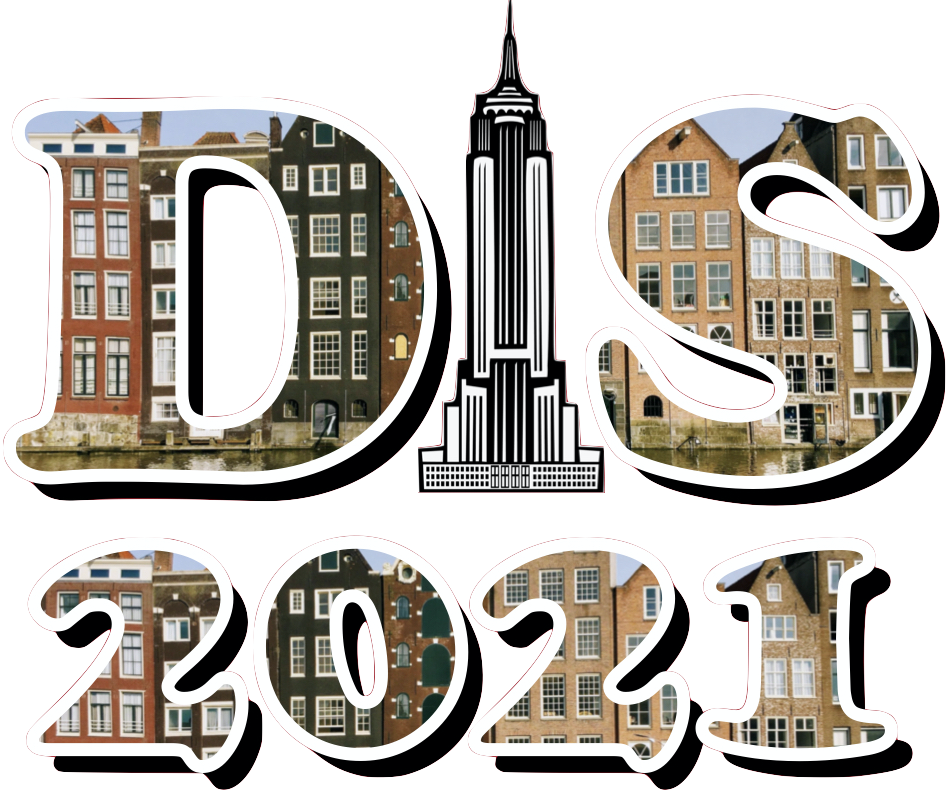}
  \end{minipage}
  &
  \begin{minipage}{0.75\textwidth}
    \begin{center}
    {\it Proceedings for the XXVIII International Workshop\\ on Deep-Inelastic Scattering and
Related Subjects,}\\
    {\it Stony Brook University, New York, USA, 12-16 April 2021} \\
    \doi{10.21468/SciPostPhysProc.?}\\
    \end{center}
  \end{minipage}
\end{tabular}
}
\end{center}

\section*{Abstract}
{\bf
Using the standard Balitsky-Fadin-Kuraev-Lipatov (BFKL) approach, with partial inclusion of next-to-leading order effects, we propose the inclusive hadroproduction of a Higgs boson and of a jet, featuring large transverse momenta and well separated in rapidity, as a new channel to probe the BFKL dynamics. Predictions are presented for cross-sections and azimuthal angle correlations in different kinematics configurations for the final-state transverse momenta. We find that the large energy scales provided by the emission of a Higgs boson stabilize the BFKL series.
}

\vspace{10pt}
\noindent\rule{\textwidth}{1pt}
\tableofcontents\thispagestyle{fancy}
\noindent\rule{\textwidth}{1pt}
\vspace{10pt}

\section{Introduction}
\label{sec:intro}
The Balitsky-Fadin-Kuraev-Lipatov (BFKL)~\cite{Fadin:1975cb,kuraev1976multi,Kuraev:1977fs,Balitsky:1978ic} approach  represents a suitable framework for the theoretical description of the QCD dynamics in the high energy-limit. During the last years, the investigation of semi-hard processes~\cite{Gribov:1983ivg} to probe the BFKL dynamics has become a theoretical and experimental challenge. Typical BFKL observables at the LHC are the azimuthal-angle correlations of the tagged particles in the final state, which are separated in rapidity, here the experimental challenge being a good resolution in the azimuthal plane, while the theoretical challenge is the incorporation of NLO corrections to impact factors, so as to treat different processes with consistent accuracy, and make predictions to be compared with data. Recently, a number of probes for BFKL signals have been proposed for different collider environments: the diffractive leptoproduction of two light vector
mesons~\cite{Ivanov_2004,Ivanov_2006,Ivanov_2007,Enberg:2005eq},
the total cross section of two highly-virtual photons~\cite{Ivanov:2014hpa},
the inclusive hadroproduction of two jets with large transverse momenta
and well separated in rapidity (Mueller-Navelet channel~\cite{Mueller:1986ey}),
for which several phenomenological studies have carried out so
far (for more details see~\cite{Celiberto:2020wpk} and references therein),
the inclusive detection of two light-charged hadrons~\cite{Celiberto_2016,Celiberto_2017,Celiberto_2017s}, three- and four-jet
hadroproduction~\cite{Caporale:2015vya,Caporale:2015int,Caporale:2016soq,Caporale:2016xku,Caporale:2016lnh,Caporale:2016zkc}, $J/\Psi$-jet~\cite{Boussarie_2018},
hadron-jet~\cite{Bolognino_2018,bolognino2019inclusive,bolognino2019highenergy}, the inclusive production of rapidity-separated $\Lambda$-$\Lambda$ or $\Lambda$-jet pairs~\cite{PhysRevD.102.094019}, and recently, double $\Lambda_{c}$ or of a $\Lambda_{c}$ plus a light-flavored jet system~\cite{celiberto2021high}, Drell-Yan--jet~\cite{Golec_Biernat_2018,Deak_2019} and heavy-quark pair
photo-~\cite{Celiberto_2018,DafneBolognino:2019ccd} and
hadroproduction~\cite{Bolognino:2019yls,Bolognino_2021}.

In this work the inclusive production at the LHC of a Higgs boson and of a jet, well separated in rapidity, is suggested as a further probe of the BFKL resummation\cite{Celiberto:2020tmb}. For a Higgs boson with mass $M_{H}=125$ GeV, the fraction of the longitudinal momentum of the parent proton carried by the struck gluon $x \sim M_{H}/\sqrt{s} \sim 0.008$ is rather small, making it describable within the BFKL approach.

\section{Theoretical Set Up}
\label{sec:theory}
\begin{figure}[h]
\centering
\includegraphics[width=0.3\textwidth]{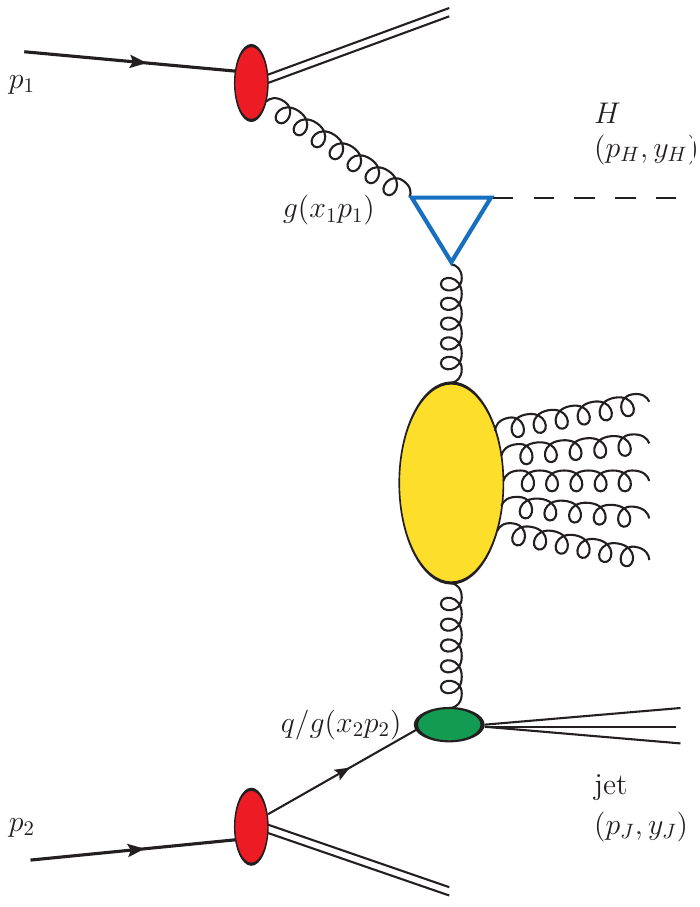}
\caption{Higgs-jet hadroproduction process}
\label{fig:process}
\end{figure}

The process of our consideration is the concurrent inclusive production of a Higgs boson
and a jet (see Fig.~\ref{fig:process}):
\begin{equation}
\label{process}
{\rm proton}(p_1) \ + \ {\rm proton}(p_2) \ \to \ H(\vec p_H, y_H) \ + \ {\rm X} \ + \ {\rm jet}(\vec p_J, y_J) \;,
\end{equation}
emitted with large transverse momenta, $|\vec p_{H,J}| \gg \Lambda_{\rm QCD}$, and
separated by a large rapidity interval, $\Delta Y = y_H - y_J$, while $p_1$ and $p_2$ are taken as
Sudakov light cone base vectors. The cross section of the process can be presented as a Fourier series of the so-called azimuthal coefficients, and it reads 
\begin{equation}
\frac{d\sigma}
{dy_H dy_J\, d|\vec p_H| \, d|\vec p_J|d\varphi_H d\varphi_J}
=\frac{1}{(2\pi)^2}\left[{\cal C}_0 + \sum_{n=1}^\infty  2\cos (n\varphi )\,
{\cal C}_n\right]\; ,
\end{equation}
where $\varphi=\varphi_H-\varphi_J-\pi$, with $\varphi_{H,J}$ the Higgs and the
jet azimuthal angles, and $C_0$ gives the total cross section, while the
coefficients $C_{n>0}$ determine their azimuthal-angle distribution. 

\section{Numerical results}
\label{sec:results}
\begin{figure}[h] 
   \centering
	\begin{minipage}[b]{0.75\linewidth}
	\hspace{-2.5cm}
		\centering
		\includegraphics[width=.55\linewidth]{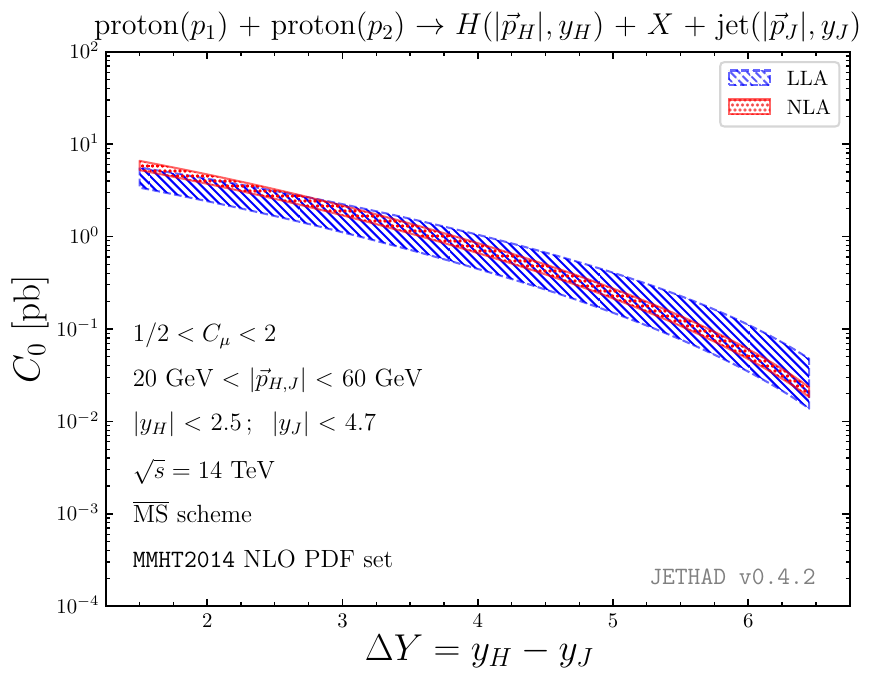} 
	\end{minipage}
	\begin{minipage}[b]{0.75\linewidth}
	 \hspace{-3.5cm}
		\includegraphics[width=.55\linewidth]{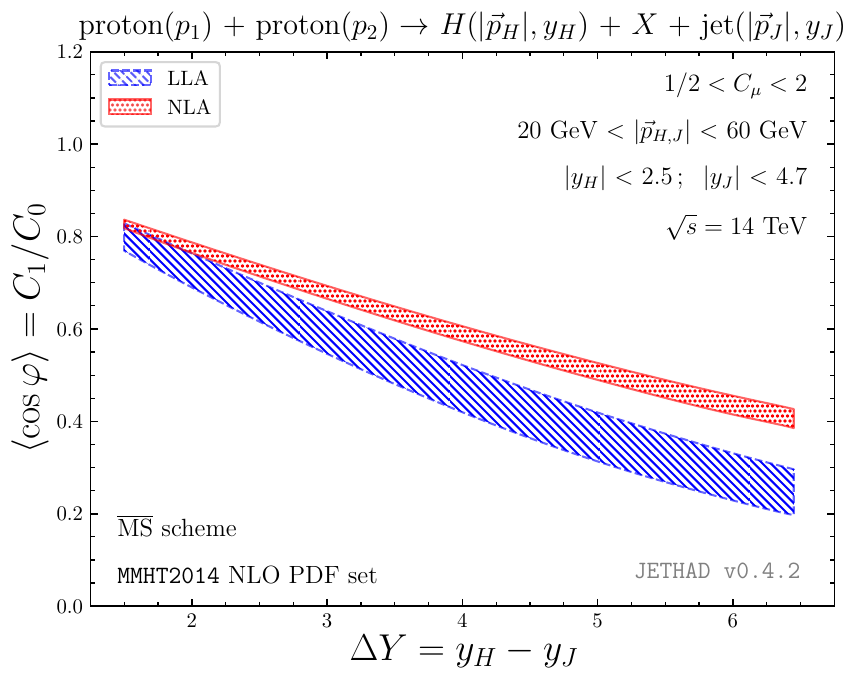}  
	\end{minipage} 
	\begin{minipage}[b]{0.75\linewidth}
		\hspace{-2.5cm}
		\centering
		\includegraphics[width=.55\linewidth]{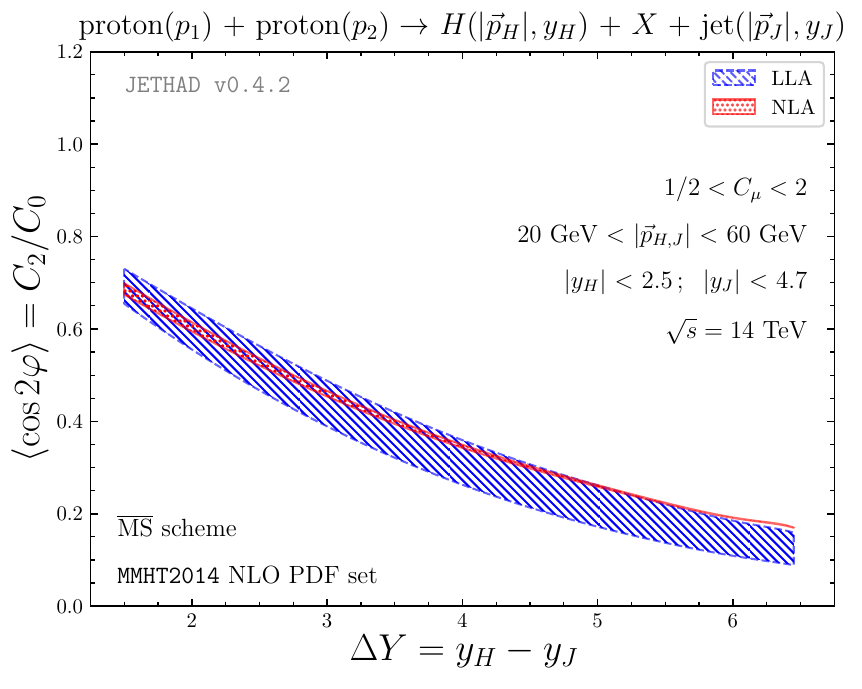} 
	\end{minipage}
	\begin{minipage}[b]{0.75\linewidth}
	 \hspace{-3.5cm}
		\includegraphics[width=.55\linewidth]{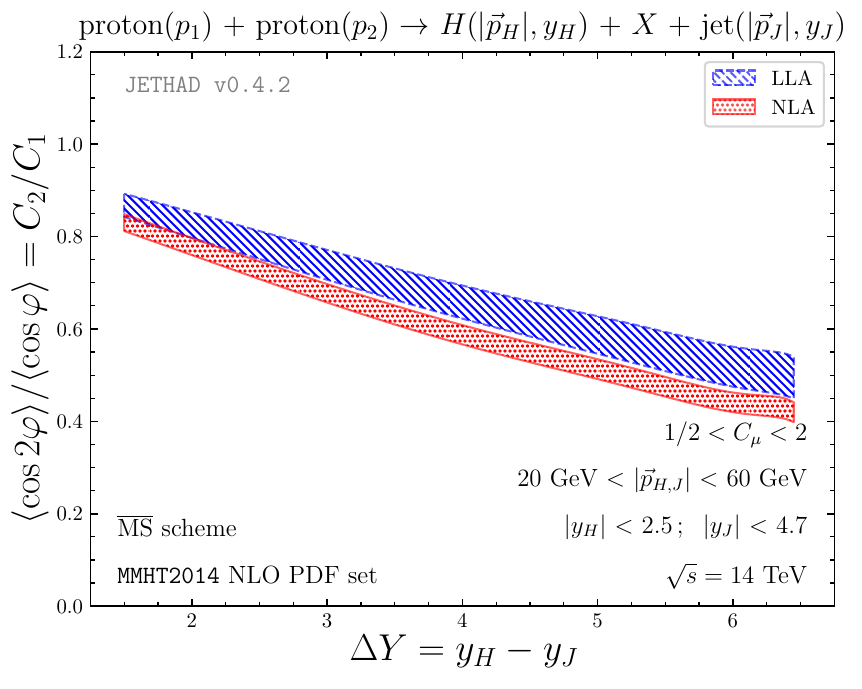}  
	\end{minipage}
	\caption{$\Delta Y$-dependence of the $C_0$ and several ratios $R_nm \equiv C_n/C_m$, for the inclusive Higgs-jet hadroproduction}
    \label{fig:c0Rnm} 
\end{figure}

\begin{figure}[h]
	\centering
	\includegraphics[width=0.85\textwidth]{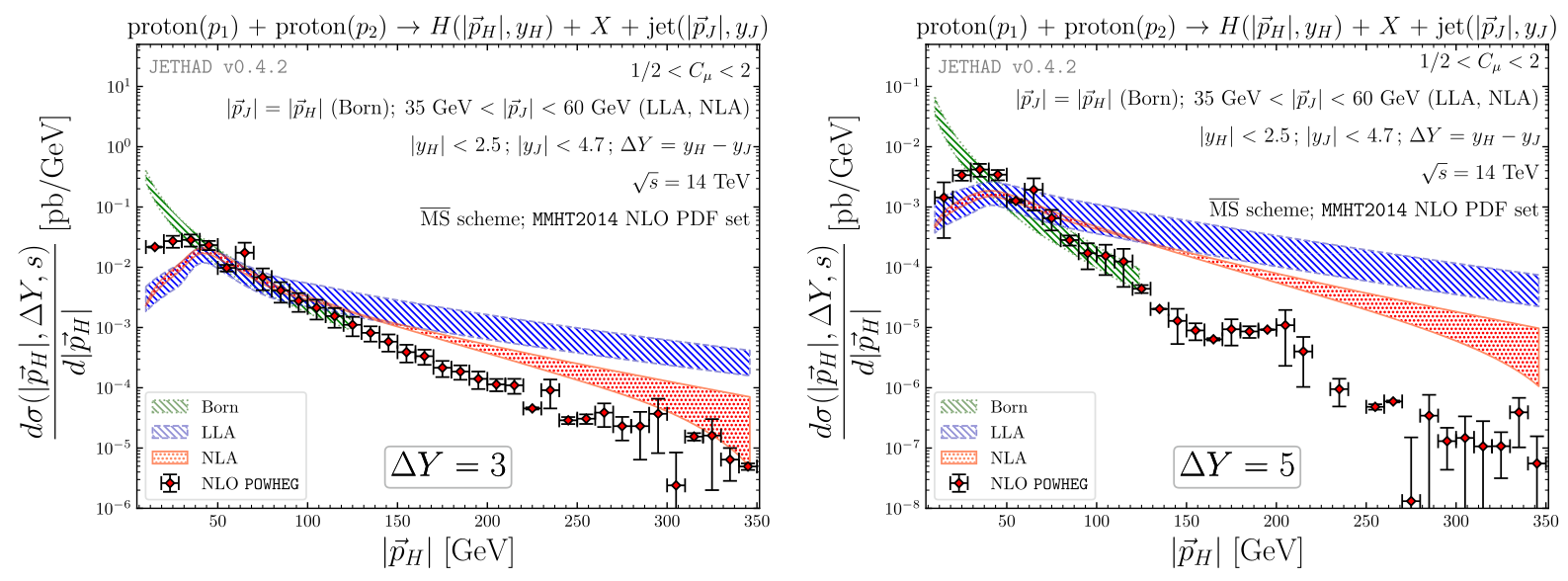}
	\caption{$p_H$-dependence of the cross section for the inclusive Higgs-jet hadroproduction}
	\label{fig:pH}
\end{figure}

The numerical analysis was preformed using \textsc{JETHAD}~\cite{Celiberto:2020wpk}, a promising standard software under development in our group, suited for the analysis of inclusive semi-hard reactions. As for the renormalization For quarks and gluon PDFs, the \textsc{MMHT2014} NLO PDF set~\cite{Harland-Lang:2014zoa} was employed. We considered three different kinematical configurations for the final-state transverse momentum of detected particles, and constrained the Higgs and jet inside the rapidity acceptances of CMS detector, $|y_H | < 2.5$ and $|y_J | < 4.7$, respectively. First, we studied the $\varphi$-summed cross section $C_0(\Delta Y,s)$, the azimuthal-correlation moments, $R_{n0}= C_n/C_0 \equiv \cos(n\phi)$, and their ratios, $ R_{nm} = C_n /C_m$ as functions of the Higgs-jet rapidity distance $\Delta Y$. We considered the $p_H$-distribution for two different values of rapidity interval ($\Delta Y = 3,5$). A detailed study on this observable covering all the high $p_H$ regions would rely on a unified formalism where distinct resummations are concurrently embodied.
We summarized our results in Figs.~\ref{fig:c0Rnm} and~\ref{fig:pH}. We adopted the MSbar renormalization scheme, obtaining, for all the considered observables, that the NLA patterns are close to LLA ones. This indicates good stability of the perturbative series, with no need to use scale optimization procedures as for other semihard processes.

\section{Conclusions}
In conclusion, the Higss-plus-jet process shows a fair stability under higher order corrections. The high energy resummation approach is valid and available in the symmetric and intermediate regions of the $p_H$ values, beyond that, the description for Higgs momentum distribution would be relied on many-sided resummation formalism unifying different approaches.

\section*{Acknowledgements}
We thank V. Bertone, G. Bozzi and L. Motyka for helpful discussions and F. Piccinini,
C. Del Pio for help on the use of the POWHEG code.


\bibliography{References}

\nolinenumbers

\end{document}